\documentclass[conference]{IEEEtran}
\IEEEoverridecommandlockouts
\usepackage{amsmath,amsfonts}
\usepackage{algorithmic}
\usepackage{algorithm}
\usepackage{array}
\usepackage[caption=false,font=footnotesize,labelfont=sf,textfont=sf]{subfig}
\usepackage{textcomp}
\usepackage{stfloats}
\usepackage{url}
\usepackage{verbatim}
\usepackage{graphicx}
\usepackage{cite}
\usepackage{color}
\def\BibTeX{{\rm B\kern-.05em{\sc i\kern-.025em b}\kern-.08em
		T\kern-.1667em\lower.7ex\hbox{E}\kern-.125emX}}
\begin{document}

\title{mmTracking: Trajectory Tracking for Uplink mmWave Devices with Multi-Path Doppler Difference of Arrival\\
\author{Cheng Lin, Chao Yu, Xiaowei Xu, Rui Wang}
}
\maketitle

\begin{abstract}
 This paper presents a method, namely mmTracking, for device trajectory tracking in a millimeter wave (mmWave) communication system. In mmTracking, the base station (BS) relies on one line-of-sight (LoS) path and at least two non-line-of-sight (NLoS) paths, which are reflected off two walls respectively, of the uplink channel to track the location of a mobile device versus time. There are at least three radio frequency (RF) chains at the BS. Analog phased array with narrow and adjustable receive beam is connected to each RF chain to capture one signal path, where the angle of arrival (AoA) can be roughly estimated. Due to the carrier frequency offset between the transmitter and the BS, the Doppler frequency of each path could hardly be estimated accurately. Instead, the differences of Doppler frequencies of the three paths can be estimated with much better accuracy. Therefore, a trajectory tracking method based on the Doppler difference and AoA estimations is proposed in mmTracking. Experimental results in a typical indoor environment  demonstrate that the average error of transmitter localization and trajectory tracking is less than 20 cm.
\end{abstract}


\section{Introduction}
\label{sec:1}

Millimeter wave (mmWave) communications have attracted significant research interests for its potential to support high data rates and low latency. However, the mmWave communication quality is sensitive to the beam misalignment or link blockage. Hence, it is necessary to exploit the great sensing potential of mmWave signals, such that the above issues could be predicted or mitigated. For example, the mmWave communication signals can be exploited in indoor layout detection \cite{sun2022indoor} and human trajectory tracking \cite{wei2015mtrack}. These results could be used to predict link blockage and prepare backup beams with static transmitter and receiver. In this paper, we would continue to show that the trajectory of mobile transmitter can also be tracked in mmWave communication systems by exploiting the multi-path channel knowledge, improving the robustness of mmWave links.

There have been a number of research efforts on the trajectory tracking of mobile devices in wireless communication systems, particularly wireless fidelity (WiFi) system. Since the time of flight (ToF) might be difficult to measure, a number of existing methods relied on time difference of arrival (TDoA) or frequency difference of arrival (FDoA). For instance, it was proposed in \cite{ho2004accurate} and \cite{musicki2009mobile} to track a mobile WiFi transmitter according to TDoA and FDoA measurements at multiple synchronized receivers. Moreover, hybrid tracking methods integrating TDoA and AoA measurements were proposed in \cite{sun2020eigenspace} and \cite{nyantakyi2024acga}. In \cite{li2023joint}, the phase difference of arrival (PDoA), TDoA and FDoA were jointly exploited to improve the target localization accuracy, where the PDoA could provide the angular information of the transmitter.
However, all these works relied on the measurements at multiple receivers, whose locations were already known and received signals were synchronized. Moreover, the measurement of TDoA at multiple receivers may be seriously distorted by the NLoS environment, which is especially the case in indoor WiFi communication. These might limit the application of the above methods in practical wireless communication systems.

There have also been a number of works on the device localization via the received signal strength indicator (RSSI) fingerprinting. For example, the survey \cite{singh2021machine} summarized a wide range of machine learning based indoor localization approaches using WiFi RSSI fingerprints. Moreover, an improved fingerprint-based localization algorithm that aggregates RSSI data from multiple access points (APs) was proposed in \cite{alfakih2020improved}. However, the accuracy of RSSI-based methods could be significantly degraded by signal fluctuations and interference. Moreover, the overhead of RSSI measurement is also significant. Finally, neither the TDoA/FDoA/PDoA-based methods nor fingerprint-based methods were demonstrated for mmWave communication systems.

In this paper, we would like to show that by exploiting superior angular resolution, a mmWave communication system could localize and track its mobile devices with single receiver. Particularly, the proposed tracking method, namely mmTracking, relies on the line-of-sight (LoS) path and two non-line-of-sight (NLoS) paths from mobile transmitter to the BS. There are at least three radio frequency (RF) chains at the BS, each with a phased array to capture the uplink signal from the desired direction. To avoid the interference of carrier frequency offset (CFO), the multi-path Doppler difference of arrival,   which is referred to as MDDoA, between NLoS path and LoS path is used, such that the CFO of different paths can be cancelled. Moreover, the AoA of LoS path can also be estimated due to the phased array. As a result, the initial location and trajectory of the mobile transmitter can be estimated according to the AoA of LoS path and the MDDoA. It is shown by experiment that the average tracking error of the proposed method is below 0.2 meters.

The remainder of this paper is organized as follows. An overview of the system architecture is provided in Section \ref{sec:2}. The signal model and the algorithm for MDDoA detection are described in Section \ref{sec:detection}. The trajectory tracking method is then elaborated in Section \ref{sec:4}. The experimental results are presented in Section \ref{sec:5}, followed by the conclusion in Section \ref{sec:6}.

\section{System Overview}
\label{sec:2}

In this paper, a novel trajectory tracking framework, namely mmTracking, is proposed for mmWave communication systems, where the single BS could track the trajectory of an uplink transmitter solely according to the MDDoA and AoA of its uplink signals. To facilitate mmTracking, the BS is equipped with at least three RF chains. Each RF chain is connected with a phased array, whose narrow receive beam can be adjusted. A uplink transmitter is moving in the service area of the BS. Its uplink signal is transmitted with a wide beam, such that both LoS and NLoS signals can be received by the BS. 

\begin{figure}[htbp]
    \centering
    \includegraphics[width=0.5\textwidth]{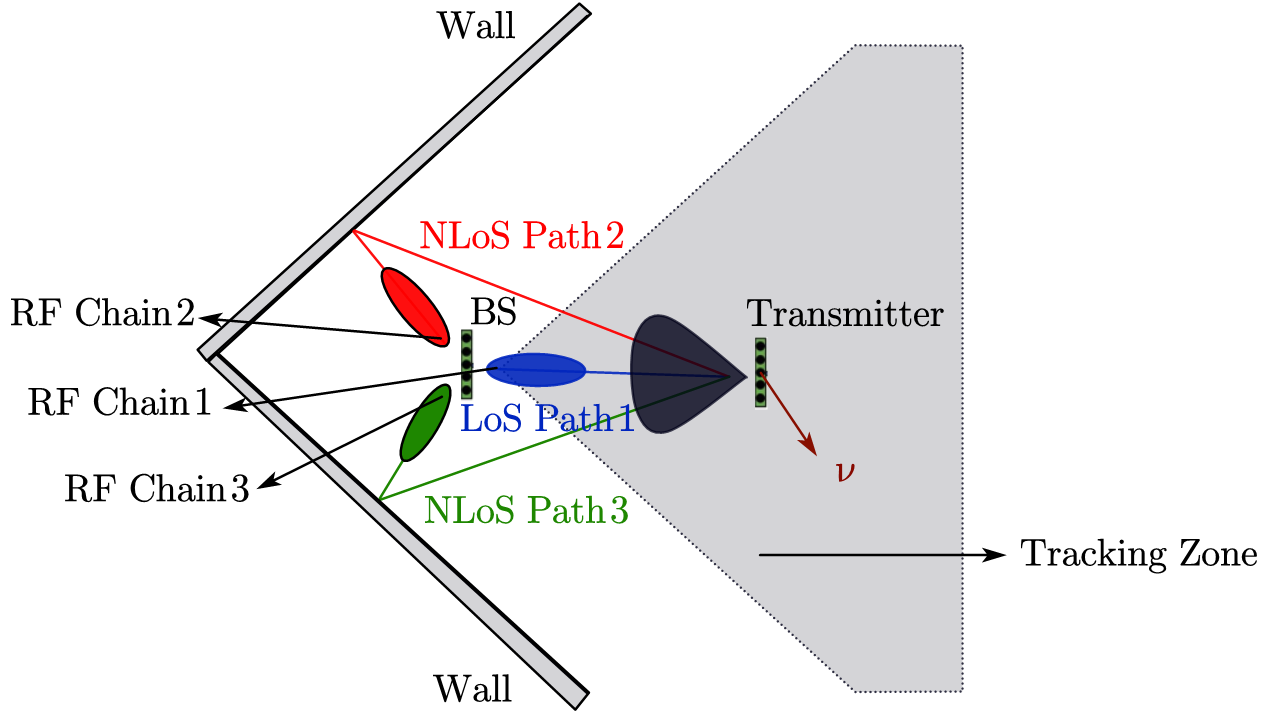}
    \caption{Example scenario of mobile transmitter tracking.}
    \label{fig: example scenario}
\end{figure}

An example scenario is illustrated in Fig. \ref{fig: example scenario}. The mmTracking framework works when there are at least two reflecting surfaces (walls) in the channel, e.g., indoor communications. It is assumed that the layout of two reflecting surfaces has been detected by the BS via the existing methods \cite{sun2022indoor,wei2017facilitating}. Hence, by adjusting the three receive beams at the BS, the signal from a mobile transmitter can be captured by the three RF chains of the BS along the LoS path and two NLoS paths via the two reflecting surfaces, respectively. 

Despite of the carrier frequency offset (CFO) between the transmitter and the BS, the BS can still detect the differences of Doppler shifts between two NLoS paths and the LoS path, respectively. For the elaboration convenience, we shall refer to the difference of Doppler shifts as the Doppler difference, and the above two Doppler differences as multi-path Doppler difference of arrival (MDDoA). Moreover, because of the phased arrays adopted at the RF chains of the BS, the AoA of each path can also be roughly detected. Due to the significant reflection loss of mmWave signal, we only adopt the AoA of LoS path in the trajectory tracking. Hence, integrating the detection on the Doppler differences and the LoS AoAs in a period, the corresponding trajectory of the mobile transmitter can be recovered. As a remark notice that the mmTracking framework does not rely on the estimation of transmitter-BS distance, which usually raises a stringent requirement on the carrier frequency synchronization between the transmitter and the BS \cite{shastri2022review}.

\section{Doppler Difference Detection}
\label{sec:detection}

In this section, the uplink signal model is first elaborated, followed by the detection of Doppler difference via the three RF chains at the BS.

\subsection{Signal Model}

\begin{figure}[htbp]
    \centering
    \includegraphics[width=0.45\textwidth]{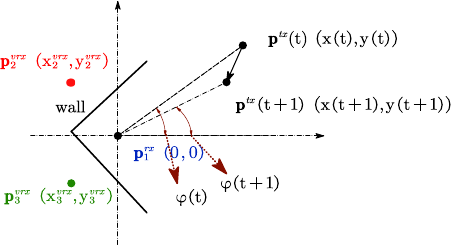}
    \caption{Illustration of the motion model in the coordinate system.}
    \label{fig: motion model}
\end{figure}

As shown in Fig. \ref{fig: motion model}, the coordinates of BS are denoted as $\mathbf{p}^{rx}_1 = (0,0)$ without loss of generality. It was shown in \cite{khawaja2020coverage} and \cite{xie2024wall} that the specular reflection is dominant off a smooth surface in mmWave band. The coordinates of BS images with respect to the two reflecting surfaces, namely virtual BSs, are denoted as ${\mathbf{p}^{rx}_{2} = (x^{rx}_{2},y^{rx}_{2})}$ and ${\mathbf{p}^{rx}_{3} = (x^{rx}_{3},y^{rx}_{3})}$, respectively. Moreover, the arbitrary trajectory of the mobile transmitter is denoted as  ${\mathbf{p}^{tx}(t) = (x(t),y(t))}$. As a remark note that the initial position of the transmitter $(x(0),y(0))$ is unknown to the BS.

The uplink signals via the LoS path and two reflection (NLoS) paths are captured by the BS, respectively. Denote the LoS path and the two NLoS paths as the first, second and third paths, respectively. Let $h_i(t)$, $\tau_i(t)$ and $f_i(t)$ be the amplitude gain, delay and Doppler frequency of the $i$-th path respectively; $\phi_{i,0}$ be the phase offset of the $i$-th path at the initial location $(x(0),y(0))$; $s(t)$ with a duration of $[0,T]$ be the information-bearing signal generated at the transmitter. Let $n_{i}(t)$ represent the interference and noise of the $i$-th path, and $\phi_o(t)$ represent the phase shift due to the CFO between the transmitter and the BS. The received signal at the $i$-th path ($i = 1, 2, 3$) can be written as
\begin{equation}
y_i(t) = h_i(t)s\bigg(t - \tau_i(t)\bigg)e^{-j\left[\phi_{i,0} + \phi_o(t) - \int_0^t 2\pi f_i(\tau) d\tau\right]} + n_i(t),
\end{equation}
 where $\phi_o(t)$ is consistent in the three RF chains of the BS. This is because the RF chains at the BS share the same clock.

The received signals of all three paths are sampled at the baseband with period $T_s$. Let $ n = 0, 1, 2, \dots, T/T_s-1$ be the sample index, the sampled signal can be written as
\begin{equation}
y_i[n] = y_i(nT_s), \ i=1,2,3.
\label{eq: discrete}
\end{equation}

\subsection{Detection of Doppler difference}

The Doppler differences are detected every $T_d$ seconds via the cross ambiguity function (CAF). We shall refer to the time instances $t=k T_d$, $k=0,1,2,...$, when the Doppler differences are detected, as the detection time instances. At each detection time instance, a window of $N_w$ baseband samples is used to calculate the CAF. 

In narrowband scenario, the propagation delay $\tau_i(t)$ can be neglected. Moreover, it is assumed that the path gain and the Doppler frequency are quasi-static within each detection window. Let $s[m]=s(m T_s)$ and $n_i[m]=n_i(mT_s)$ be the sampled transmit signal and noise in baseband respectively, $N_0=T_d/T_s$, $f_{i,k}=f_i(kT_d)$, the sampled baseband signal in the window of the $k$-th detection time instance can be represented as
\begin{equation}
\begin{split}
    y_i[m] &= h_{i}(kT_d)s[m]e^{-j[\phi_{i,0}+\phi_{o}(mT_s)-\phi_{f,m}]} + n_i[m],\\
    &\quad kN_0\leq m \leq kN_0 + N_w -1,
\end{split} \label{eqn:ym}
\end{equation}
where 
\begin{equation}
    \phi_{f,m} = \int_0^{kT_d} 2\pi f_i(\tau) d\tau + f_{i,k}(m-kN_0)T_s \label{eqn:doppler-phase}
\end{equation}
represents the accumulated phase shift due to Doppler frequency. Then, the CAF between the $i$-th path ($i=2,3$) and the LoS path is defined as
\begin{equation}
R_{i,k}(f) = \sum_{n=kN_0}^{kN_0 + N_w - 1} {y}_{i}[n] y^{*}_{1}[n] e^{-j2\pi f (n-kN_0) T_s},
\end{equation}
where $(.)^*$ is the complex conjugate. 

Define the Doppler difference of the $k$-th detection time instance as $$f_{i,k}^d = f_{i,k}-f_{1,k}.$$ Since $\phi_o(mT_s)$ in (\ref{eqn:ym}) is consistent in the three RF chains, a peak value of $R_{i,k}$ can be found at $f=f_{i,k}^d$. Thus, the Doppler difference between the $i$-th path and the LoS path can be detected via the above CAF. However, there might be multiple peaks in the CAF due to interference. To minimize false detection, an adaptive threshold-based method is applied to extract the Doppler difference from the CAF. Particularly, the Doppler difference $f$ is detected if
\begin{align*}
    R_{i,k}(f) \geq \beta_{i,k}(f),
\end{align*}
where $\beta_{i,k}(f)$ is the detection threshold. It can be calculated as
\begin{equation}
\beta_{i,k}(f) = \frac{\gamma}{2W + 1} \sum_{p=-W}^{W} R_{i,k}(f + p\Delta f),
\label{eq: beta}
\end{equation}
where $W$ is the half length of training cells, ${\gamma>1}$ is a scaling factor for the detection threshold, $\Delta f = \frac{1}{N_wT_s}$ is the resolution of the Doppler difference. In the case that there are still multiple detected Doppler differences after the above threshold-based filtering, the one with the maximum value of $R_{i,k}$ is selected as the detected Doppler difference. In the remaining of this paper, we shall refer to the detected Doppler difference at the $k$-th detection time instance as $\hat{f}_{i,k}^d$, $i=2,3$.

\section{Trajectory Tracking}
\label{sec:4}

The mmTracking framework relies on the first two detection time instances to estimate the initial position of the transmitter, and track its trajectory according to the accumulated phase offset due to MDDoA and the AoA of LoS path.  In the following parts, we first establish the geometric model of motion, and then elaborate the method of trajectory tracking.

\subsection{Motion Model}

The Doppler frequency of the LoS path is due to the change of BS-transmitter distance, and the Doppler frequencies of the two NLoS paths are equivalently due to the distance variation between the virtual BSs and the transmitter respectively. Hence, we have
\begin{equation}
\frac{\partial  \| \mathbf{p}^{rx}_i - \mathbf{p}^{tx}(t) \| }{\partial t} = - \lambda \cdot{f}_{i}(t), i = 1,2,3,
\end{equation}
where $\lambda$ represents the wavelength of the carrier frequency. Moreover, define the distance difference between LoS path and $i$-th path ($i=2,3$) at time instance $t$ as
\begin{equation}
d_{i}(t) 
=  \| \mathbf{p}^{rx}_i - \mathbf{p}^{tx}(t) \| -  \| \mathbf{p}^{rx}_1 - \mathbf{p}^{tx}(t) \|,
\label{eq: distance difference}
\end{equation}
we have
\begin{equation}
\frac{\partial\ d_{i}(t)}{\partial t} = -\lambda \cdot \bigg(f_{i}(t)-f_{1}(t)\bigg), i=2,3.
\label{eq: difference}
\end{equation}
Particularly,
\begin{equation}
\frac{\partial\ d_{i}(t)}{\partial t}\bigg|_{t=kT_d} = -\lambda \cdot f^d_{i,k}, i=2,3.
\label{eq: differenceK}
\end{equation}
Then, assuming the Doppler difference is quasi-static within the duration of $T_d$, we have
\begin{equation}
d_{i}((k+1)T_d) = d_{i}(kT_d) - \lambda \cdot f^d_{i,k} \cdot T_d .
\label{eq: distance increase}
\end{equation}

Moreover, let $\varphi_k$ be the AoA of LoS path at $k$-th time detection time instance, we have 
\begin{equation}
\varphi_k = \arctan \left(\frac{y(k T_d)}{x(k T_d)}\right).
\label{eq: AoA}
\end{equation}

\subsection{Initial Position Detection}

As discussed in the previous section, at each detection time instance (say the $k$-th one), the Doppler differences $f_{2,k}^d, f_{3,k}^d$ and AoA of LoS path $\varphi_k$ can be estimated by mmTracking, denoted as $\hat{f}_{2,k}^d, \hat{f}_{3,k}^d$, and $\hat{\varphi}_k$ respectively.
In this part, the positions of first two detection time instances, denoted as $\mathbf{p}_0=(x(0),y(0))$ and $\mathbf{p}_1=(x(T_d),y(T_d))$, are jointly detected via the Doppler differences $\{\hat{f}^d_{2,0},\hat{f}^d_{3,0}\}$ and AoA  $\{\hat{\varphi}_0,\hat{\varphi}_1\}$. 

Particularly, the estimations of $\mathbf{p}_0$ and $\mathbf{p}_1$ minimizing the weighted mean square error (MSE) of Doppler differences and LoS AoA will be found. From (\ref{eq: distance increase}), the estimation error of $f_{i,0}^d$ ($i=2,3$) can be written as
\begin{align}
e^f_{i} &= f_{i,0}^d - \hat{f}_{i,0}^d \nonumber \\
&= \frac{1}{\lambda T_d} \left( d_i(0) - d_i(T_d) \right) - \hat{f}_{i,0}^d \nonumber \\
&= \frac{1}{\lambda T_d} \Bigg[ \left( \| \mathbf{p}^{rx}_i - \mathbf{p}_0 \| - \| \mathbf{p}^{rx}_1 - \mathbf{p}_0 \| \right) \nonumber \\
&\quad - \left( \| \mathbf{p}^{rx}_i - \mathbf{p}_1 \| - \| \mathbf{p}^{rx}_1 - \mathbf{p}_1 \| \right) \Bigg] - \hat{f}_{i,0}^d.
\end{align}
Moreover, the estimation errors of $\varphi_0$ and $\varphi_1$ are given by
\begin{equation}
e^{a}_k = \varphi_k - \hat{\varphi}_k = \arctan \left( \frac{y(kT_d)}{x(kT_d)} \right) - \hat{\varphi}_k, \quad k = 0, 1.
\end{equation}

Hence, the vector aggregating the estimation errors can be expressed as $$\mathbf{e}=(e^f_{2},e^f_{3},e^{a}_0,e^{a}_1)^{\top},$$ where $(\cdot)^{\top}$ represents the matrix transpose. The weighted MSE of Doppler difference and LoS AoA can be expressed as
\begin{equation}
    g(\mathbf{p}) = \mathbf{e}^{\top} \mathbf{W} \mathbf{e},
    \label{eqn:g(x)}
\end{equation}
where $\mathbf{W} = \mathrm{diag}(w_1, w_2, w_3, w_4)$ is the matrix of weights, and $\mathbf{p} = (\mathbf{p}_0, \mathbf{p}_1) = (x(0), y(0), x(T_d), y(T_d))$ for notation convenience. The detection of initial positions $\mathbf{p}$ can be formulated as the following optimization problem: 
\begin{equation}
\begin{aligned}
\hat{\mathbf{p}} = (\hat{\mathbf{p}}_0, \hat{\mathbf{p}}_1) = \arg\min_{\mathbf{p}} g(\mathbf{p}).
\end{aligned}
\label{eqn:localization}
\end{equation}

The above problem can be efficiently solved using the Newton-Raphson method as elaborated in \cite{pho2022improvements}, where the solution can be found iteratively. Since the objective is differentiable respect to all variables in $\mathbf{p}$, the Newton-Raphson update in the $i$-th iteration is written as 
\begin{align}
  \mathbf{p}^{(i+1)} = \mathbf{p}^{(i)} - \mathbf{H}^{-1} \nabla_{\mathbf{p}} g(\mathbf{p}),  
\end{align}
where the gradient of the objective is
\[
\nabla_{\mathbf{p}} g(\mathbf{p}) = 2 \mathbf{J}^{\top} \mathbf{W} \mathbf{e},
\] 
the Hessian matrix can be approximated as
\[
\mathbf{H} = 2 \mathbf{J}^{\top} \mathbf{W} \mathbf{J},
\]
and the Jacobian matrix \( \mathbf{J} \) is
\[
\mathbf{J} = 
\begin{bmatrix}
\frac{\partial e^f_2}{\partial x(0)} & \frac{\partial e^f_2}{\partial y(0)} & \frac{\partial e^f_2}{\partial x(T_d)} & \frac{\partial e^f_2}{\partial y(T_d)} \\
\frac{\partial e^f_3}{\partial x(0)} & \frac{\partial e^f_3}{\partial y(0)} & \frac{\partial e^f_3}{\partial x(T_d)} & \frac{\partial e^f_3}{\partial y(T_d)} \\
\frac{\partial e^a_0}{\partial x(0)} & \frac{\partial e^a_0}{\partial y(0)} & \frac{\partial e^a_0}{\partial x(T_d)} & \frac{\partial e^a_0}{\partial y(T_d)} \\
\frac{\partial e^a_1}{\partial x(0)} & \frac{\partial e^a_1}{\partial y(0)} & \frac{\partial e^a_1}{\partial x(T_d)} & \frac{\partial e^a_1}{\partial y(T_d)}
\end{bmatrix}.
\]

The iterative algorithm terminates when \( \|\nabla_{\mathbf{p}} g(\mathbf{p})\| < \epsilon \), where \( \epsilon \) is the convergence threshold.

\subsection{Trajectory Tracking Method}

Given the estimation of the initial position ${\hat{\mathbf{p}}_0 = (\hat{x}(0),\hat{y}(0))}$, the transmitter position at $k$-th detection time instance $\mathbf{p}_k = (x(k T_d),y(k T_d))$ can be estimated according to the accumulation of Doppler differences. Particularly, the estimation of $\mathbf{p}_k$, denoted as $\hat{\mathbf{p}}_k = (\hat{x}(k T_d),\hat{y}(k T_d))$, will be found to minimize the weighted MSE of the accumulated Doppler differences since the first detection time instance and the LoS AoA. 

From \eqref{eq: differenceK} and \eqref{eq: distance increase}, the accumulated Doppler differences can be expressed as 
\begin{equation}
    \sum_{m=0}^{k-1} {f}^d_{i,m} = \frac{1}{\lambda T_d} \bigg( d_i(0) - d_i(kT_d) \bigg).
\end{equation}
Hence, its detection error can be written as
\begin{equation}
\begin{split}
    n^f_{i,k} &= \sum_{m=0}^{k-1} {f}^d_{i,m} - \sum_{m=0}^{k-1} \hat{f}^d_{i,m} \\
    &= \frac{1}{\lambda T_d} \bigg( d_i(0) -  \| \mathbf{p}^{rx}_i - \mathbf{p}_k \| +  \| \mathbf{p}^{rx}_1 - \mathbf{p}_k \| \bigg)  \\
    &\quad  -  \sum_{m=0}^{k-1} \hat{f}^d_{i,m}  , 
\end{split}
\end{equation}
where the initial distance difference ${d}_i(0)$ between the $i$-th path and the LoS path can be calculated as
\begin{equation*}
    {d}_i(0) \approx \hat{d}_i(0) = \| \mathbf{p}^{rx}_i - \hat{\mathbf{p}}_0 \| -  \| \mathbf{p}^{rx}_1 - \hat{\mathbf{p}}_0 \|.
\end{equation*}
Moreover, the estimation error of AoA for the LoS path at the $k$-th detection time instance can be written as
\begin{equation}
    n_k^a = {\varphi}_k - \hat{\varphi}_k = \arctan \left(\frac{y(kT_d)}{x(kT_d)} \right) - \hat{\varphi}_k.
\end{equation}

Similar to the detection of initial positions, the vector aggregating the estimation errors can be expressed as $$\mathbf{n}=(n^f_{2,k},n^f_{3,k},n^{a}_k)^{\top},$$ and the weighted MSE of the accumulated Doppler differences and AoA can be expressed as 
\begin{equation}
    h(\mathbf{p}_k) = \mathbf{n}^{\top} \widetilde{\mathbf{W}} \mathbf{n},
\end{equation}
where $\widetilde{\mathbf{W}} = \mathrm{diag}(\widetilde{w}_1, \widetilde{w}_2, \widetilde{w}_3)$ is the matrix of weights. Hence, the detection of position at $k$-th detection time instance $\mathbf{p}_k$ can be formulated as the following optimization problem:
\begin{equation}
\begin{aligned}
\hat{\mathbf{p}}_k
&= \arg\min_{\mathbf{p}_k}   h(\mathbf{p}_k),
\end{aligned}
\label{eqn:tracking}
\end{equation}
which can also be solved by the Newton-Raphson method.

\section{Experiment Results and Discussion}
\label{sec:5}

\begin{figure}[htbp]
    \centering
    \includegraphics[width=0.5\textwidth]{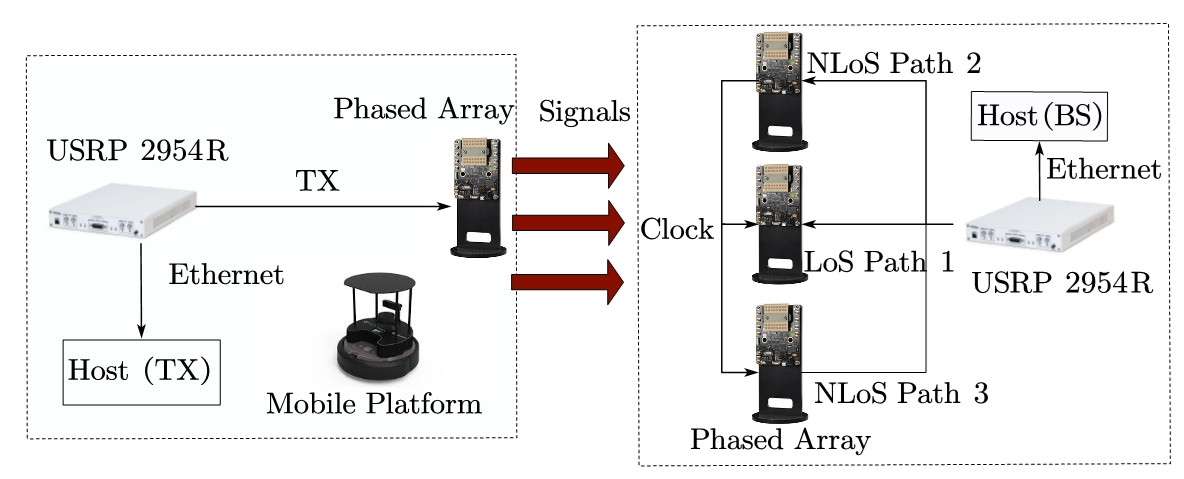}
    \caption{Diagram of system implementation.}
    \label{fig: system implemention}
\end{figure}

\begin{figure}[htbp]
    \centering
    \includegraphics[width=0.45\textwidth]{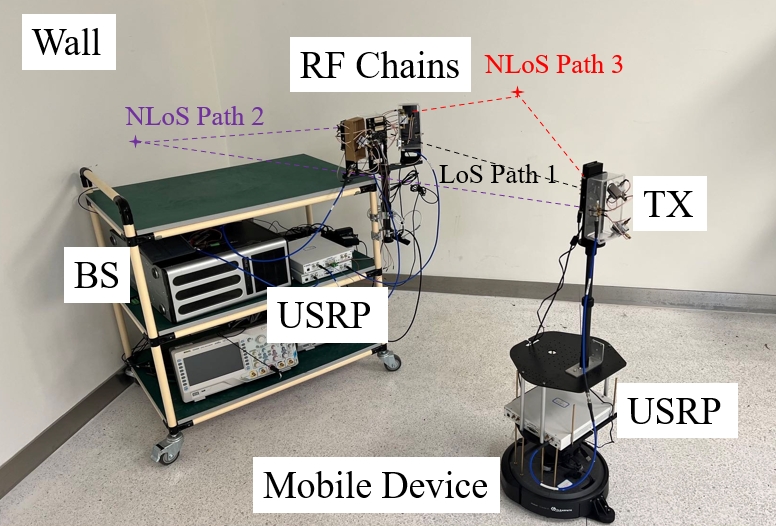}
    \caption{Illustration of the experiment environment.}
    \label{fig: experiment environment}
\end{figure}

The implementation of the mmTracking system is shown in Fig. \ref{fig: system implemention}. The transmitter is implemented by an NI USRP-2954R software-defined radio (SDR) connected to a Sivers mmWave phased array and a data processing laptop. The whole transmitter is deployed on a TurtleBot mobile robot, such that its trajectory can be controlled. The transmitted signal is modulated by the orthogonal frequency division multiplexing (OFDM) technology with a bandwidth of $1$ MHz, consisting of a training sequence and data payload. The carrier frequency of the transmitter is $60.48$ GHz, and the transmitting beam width is $90^{\circ}$. At the receiver side, three Sivers phased arrays are connected to an NI USRP-2954R to capture the signals of three paths, respectively. To ensure that the receive beams are aligned with the desired signals, each receive phased array periodically performs beam search on the potential arrival directions of desired signal. The experiment scenario is illustrated in Fig. \ref{fig: experiment environment}, where the receiver (BS) locates at the corner of two walls. The detection interval is $T_d$ = 0.05s, and the sampling frequency $1/T_s$ = 2MHz.

\begin{figure}[htbp]
    \centering
    \includegraphics[width=0.45\textwidth]{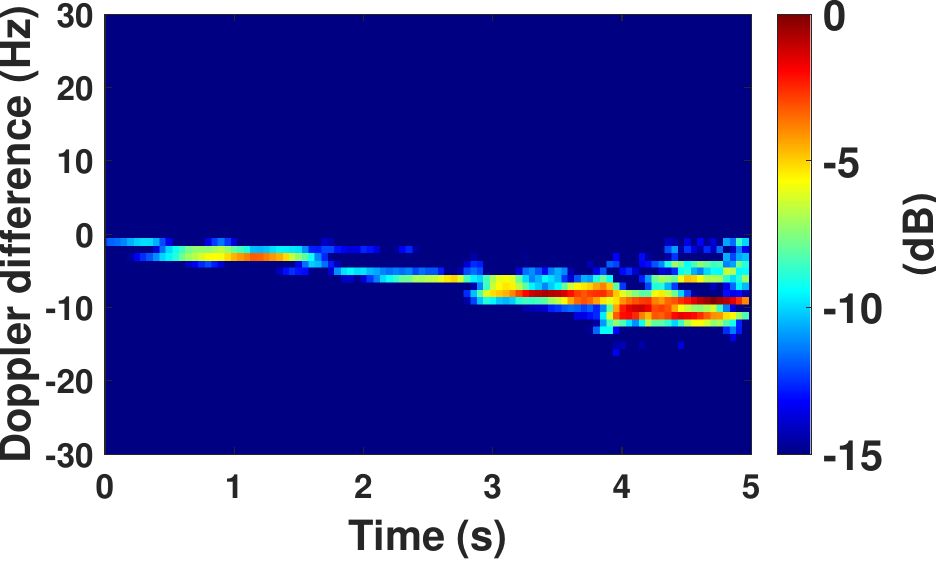}
    \caption{MDDoA between path 2 and LoS Path.}
    \label{fig: stftsubfig1}
\end{figure}

\begin{figure}[htbp]
    \centering
    \includegraphics[width=0.45\textwidth]{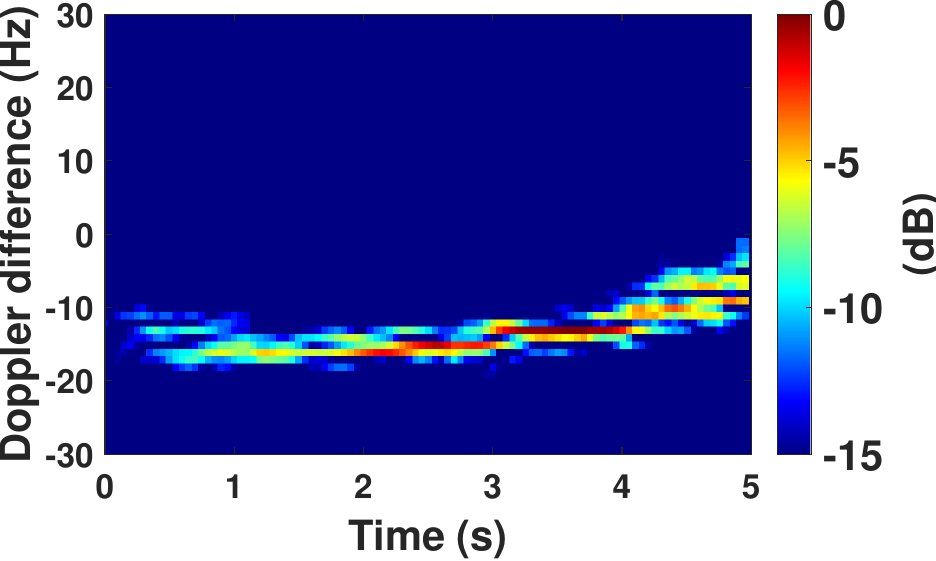}
    \caption{MDDoA between path 3 and LoS Path.}
    \label{fig: stftsubfig2}
\end{figure}

\begin{figure}[htbp]
    \centering
    \includegraphics[width=0.5\textwidth]{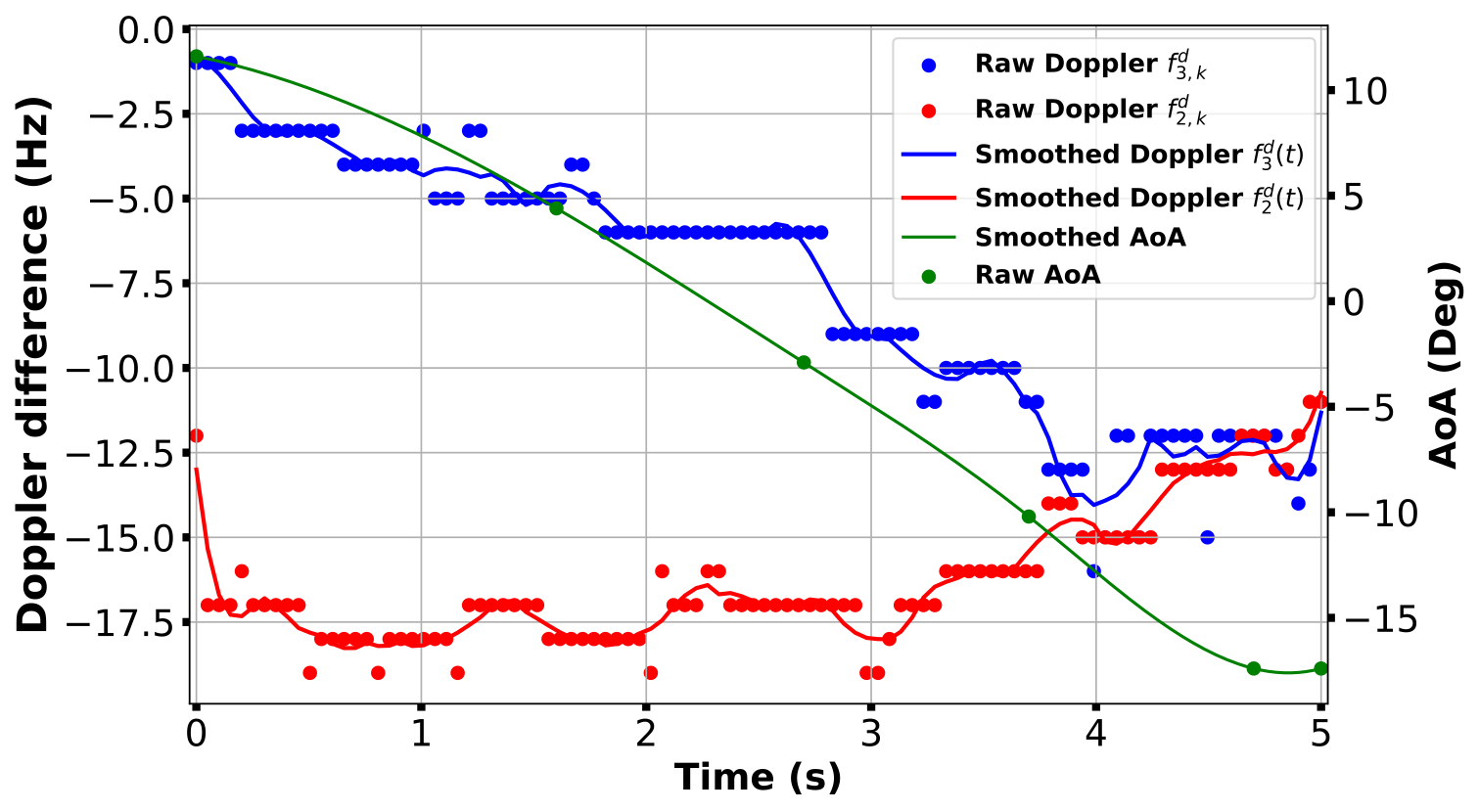}
    \caption{Estimated and smoothed Doppler differences and AoAs.}
    \label{fig: doppler+AoA}
\end{figure}

The detected Doppler difference spectrograms (Doppler differences versus time) between the two NLoS paths and the LoS path of a trajectory are illustrated in Fig. \ref{fig: stftsubfig1} and Fig. \ref{fig: stftsubfig2}, respectively. Applying the Doppler difference detection method introduced in Section \ref{sec:detection}, the AoAs of the LoS path and the Doppler differences of both NLoS paths  are both illustrated in Fig. \ref{fig: doppler+AoA}, where polynomial smoothing is applied to improve the accuracy of raw detection. 


\begin{figure}[htbp]
    \centering
    \includegraphics[width=0.45\textwidth]{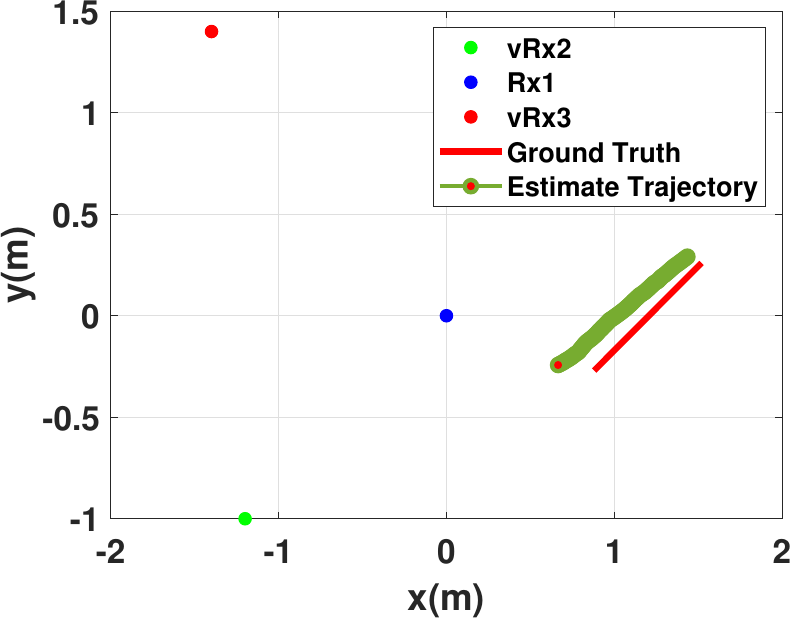}
    \caption{Tracking result of trajectory 1.}
    \label{fig: trackingsubfig1}
\end{figure}
\begin{figure}[htbp]
    \centering
    \includegraphics[width=0.45\textwidth]{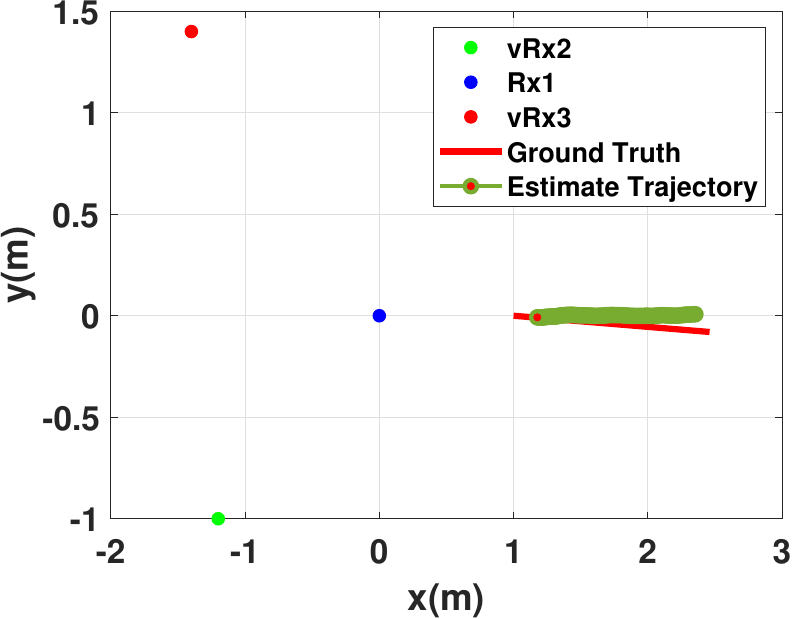}
    \caption{Tracking result of trajectory 2.}
    \label{fig: trackingsubfig2}
\end{figure}

\begin{figure}[htbp]
    \centering
    \includegraphics[width=0.42\textwidth]{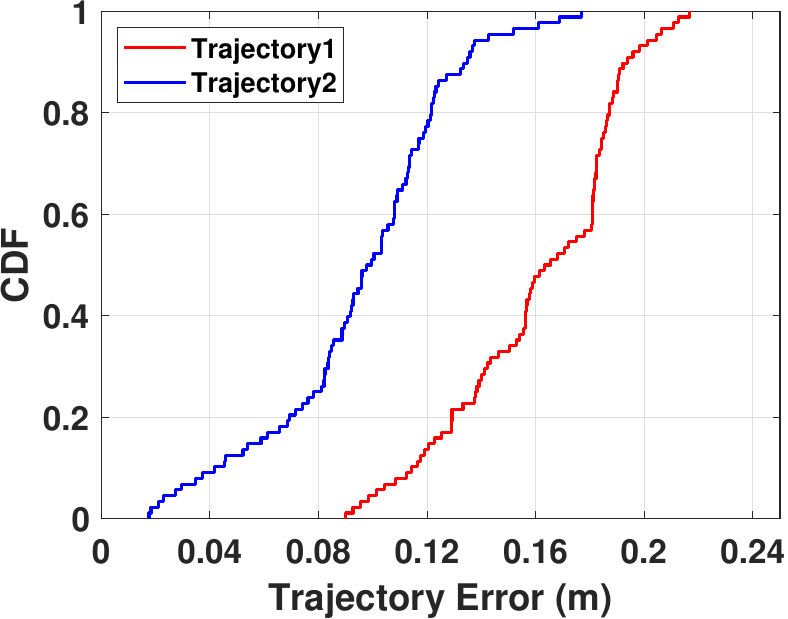}
    \caption{The CDFs of trajectory estimation error.}
    \label{fig: cdf}
\end{figure}

The comparison of the estimated trajectory and the actual one for the above detection is illustrated in Fig. \ref{fig: trackingsubfig1}, and the tracking of another trajectory is shown in Fig. \ref{fig: trackingsubfig2}. In both figures, the red curves represent the ground truth trajectories of the mobile transmitter, while the green curves denote the trajectories estimated using the proposed method. It can be observed that the localization errors of initial positions in both trials are within $0.15$m. Despite of minor fluctuations in the estimated parameters, the estimated trajectories closely follow the ground truth, demonstrating the accuracy of the proposed tracking approach. Finally, the cumulative distribution functions (CDFs) of the localization errors for two different trajectories are illustrated in Fig. \ref{fig: cdf}. It can be observed that $90\%$ estimation errors are within $0.2$m.

\section{Conclusion}
\label{sec:6}

This paper proposes a method of trajectory tracking for mobile mmWave devices via MDDoA and AoA of uplink signal. Particularly, multiple RF chains at the receiver capture multi-path signals from different beam directions periodically. These signals are then used to detect the Doppler differences between paths and the AoA of the LoS path. Based on the detection, the initial position and trajectory of the mmWave device can be estimated. Compared with the existing trajectory tracking methods for WiFi systems, the proposed method can be conducted with single receiver. Experimental results demonstrate an average tracking error below 0.2 meters.

\bibliographystyle{IEEEtran}
\bibliography{main}
\end{document}